\documentclass{emulateapj}

\usepackage{amsmath}
\usepackage{natbib}
\usepackage[colorlinks=true,citecolor=blue,urlcolor=magenta,breaklinks]{hyperref}
\usepackage{cool}
\usepackage{eulervm}

\def\simpropto{\lower.2ex\hbox{$\; \buildrel \propto \over \sim \;$}}
\def\ltsim{\lower.5ex\hbox{$\; \buildrel < \over \sim \;$}}
\def\gtsim{\lower.5ex\hbox{$\; \buildrel > \over \sim \;$}}
\begin{document}
 
\title[Helioseismology and Asteroseismology: Looking for Gravitational Waves in  acoustic oscillations]{
Helioseismology and Asteroseismology: Looking for Gravitational Waves in acoustic oscillations}
\author{Il\'\i dio Lopes~\altaffilmark{1,2} {\sc and} Joseph Silk~\altaffilmark{3,4,5}}
\altaffiltext{1}{Centro Multidisciplinar de Astrof\'{\i}sica, Instituto Superior T\'ecnico, 
Universidade de Lisboa, Av. Rovisco Pais, 1049-001 Lisboa, Portugal; ilidio.lopes@tecnico.ulisboa.pt} 
\altaffiltext{2}{
Departamento de F\'\i sica, Escola de Ciencia e Tecnologia, 
Universidade de \'Evora, Col\'egio Luis Ant\'onio Verney, 7002-554 \'Evora, Portugal; ilopes@uevora.pt} 
\altaffiltext{3}{Institut d'Astrophysique de Paris, UMR 7095 CNRS, Universit\'e Pierre et Marie Curie, 
98 bis Boulevard Arago, Paris 75014, France; silk@astro.ox.ac.uk} 
\altaffiltext{4}{Beecroft Institute of Particle Astrophysics and Cosmology, 1 Keble Road, 
University of Oxford, Oxford OX1 3RH, UK} 
\altaffiltext{5}{
Department of Physics and Astronomy, 3701 San Martin Drive, The Johns Hopkins University, Baltimore, MD 21218, USA
} 


\begin{abstract} 
Current helioseismology observations allow the determination of the frequencies and 
surface velocity amplitudes of  solar acoustic modes with exceptionally high precision. 
In some cases, the frequency accuracy is better than one part in a million. 
We show that there is a distinct possibility that the quadrupole acoustic modes of low order 
could be excited by gravitational waves (GWs), if the GWs have a strain amplitude in the range $10^{-20}h_{-20}$ 
with  $h_{-20}\sim 1$ or $h_{-20}\sim 10^{3}$, as predicted by  
several types of GW sources, such as galactic ultracompact binaries 
or  extreme mass ratio inspirals and coalescence of black holes. 
If the damping rate at low order is  $ 10^{-3}\eta_N$ $\mu{\rm Hz}$, 
with $\eta_N\sim 10^{-3} $ -- $1,$ as inferred from the theory of stellar pulsations, then
GW radiation will lead to a maximum rms surface velocity amplitude of quadrupole modes 
of the order of $h_{-20}\eta_N^{-1}\sim$ $10^{-9}$ -- $10^{-3}$ ${\rm cm\; s^{-1}}$,
on the verge of what is currently detectable via helioseismology.   
The frequency and sensitivity range probed by  helioseismological acoustic modes overlap with, and complement,   the  capabilities of eLISA for 
the brightest resolved ultracompact galactic binaries.
\end{abstract}


  
\section{Introduction}

The rapid development of gravitational wave (GW) detection by either resonant mass detectors 
or ground-based and space interferometers give us hope that  GW observations will very soon become a reality.
If such a goal were to be achieved,   a new window will open toward understanding the formation of many compact structures 
in the universe, most of which are still poorly understood, such as black holes and neutron star binaries
~\citep[e.g.,][]{2013LRR....16....7G,2009LRR....12....2S}. Nevertheless, even if the detection of GW radiation 
can be achieved by these  modern experiments, the goal will only be attained if the GW signal can be successfully  separated from the background "noise". Therefore,  any prior information of incoming GW events~\citep[e.g.,][]{2009RPPh...72g6901A,2009LRR....12....2S} for the GW experimental research community is of great interest.

In this paper, we discuss an alternative method to probing for direct GW radiation.
The Sun, as is the case for many other stars, is a natural massive GW detector with an isotropic sensitivity to  
GWs, able to absorb GWs from any direction of the sky.  
In recent years, this possibility of using stars as GW detectors
has become very appealing, as current helioseismology and asteroseismology observations 
allow the determination of the frequency and the velocity amplitude of many modes of vibrations with exceptional accuracy.

In the Sun, the acoustic modes have been continuously observed by the {\it SOHO} mission 
since 1996~\citep{2012RAA....12.1107T} and  some of the low degree modes are measured with a precision 
of one part per million.
The {\it COROT}~\citep{2008Sci...322..558M} and {\it Kepler}~\citep{2011Sci...332..213C} missions have  discovered  more than 500 pulsating stars, most of which are in the main sequence and sub giant phase, and some of these stars have been observed for priors of several months in the last 4 years~\citep[e.g.,][]{2013ARA&A..51..353C}. Like for
the Sun the damping and excitation of the oscillation modes in these stars  
is attributed to turbulent convection in their upper layers. 
The continuous monitoring of pulsating modes in the Sun and many stars of 
different masses and sizes give us the possibility of surveying the local universe 
for GW radiation, either by probing for a stochastic background, or for rare events or  for
periodic signals~\citep[e.g.,][]{2009LRR....12....2S}.  
Among other possible GW sources emitting  in the frequency range of  
solar acoustic oscillations ($0.2\;{\rm mHz} \le \nu \le 5 {\rm mHz} $),
there are the occasional GW events occurring during the coalescence of massive black hole binaries
and neutron star binaries~\citep{1971MNRAS.152..461L}, 
extreme mass ratio inspirals~\citep{2012arXiv1210.8066G},
and the 
periodic GW signal of AM CVn stellar systems~\citep{2004MNRAS.349..181N,2007ApJ...666.1174R,2010A&A...521A..85Y}.
The strain amplitude of these GW events is in the range of
$10^{-17}$ -- $10^{-24}$~\citep[e.g.,][]{2009LRR....12....2S,Moore:wk}.

Preliminary studies of the impact of incoming GW radiation on massive bodies, 
such as the Earth, Moon, planets and stars were previously presented by several authors
~\citep[e.g.,][]{1969ApJ...156..529D,1980ApJ...241..475Z,1984ApJ...286..387B,1997A&A...321.1024K}.
\citet{1984ApJ...286..387B} were the first to compute the impact of GW on solar gravity and acoustic modes,
for which they also put upper-limits on the stochastic gravitational background from the observed solar oscillations. 
More recently~\citet{2011ApJ...729..137S} use an hydrodynamical model to re-evaluate the excitation of 
solar oscillations by GWs~\citep{2010MNRAS.408.1742S}. 
Equally they have updated the previous stochastic gravitational background limits~\citep{2014ApJ...784...88S}. 
A complementary approach was performed recently~\citep{2014arXiv1405.1414M} in which the authors 
estimated that gravitational radiation that is absorbed by stars near black holes,
and discuss how the absorption by the Sun of GWs from Galactic white dwarf binaries
could be observed by a second generation of gravitational wave detectors.

Here, we show that GWs with a strain spectral amplitude of $10^{-20}h_{-20}$ with $h_{-20} \gtsim 1$
can lead to the excitation of low order quadrupole acoustic modes in the Sun, 
for which the rms surface velocity amplitudes could be as large as $\sim h_{-20}\;{\rm cm\; s^{-1}}$.
These results use theoretical predictions of  damping rates of acoustic modes consistent 
with current solar observations at high frequencies.
Moreover, we discuss  the strategy to search for GW events in stellar oscillations.
Our theoretical model closely follows the GW model of resonant mass detectors. 
This approach facilitates the use of our work by the GW experimental community.    
             
\section{Gravitational waves and stellar oscillations}

In the presence of GWs, stars behave like resonant-mass spherical detectors.
Accordingly, the oscillations of a star equally excited by convection and 
GWs can be accurately represented by the simplified wave
equation~\citep[e.g.,][]{2005MNRAS.360..859C,2001A&A...370..136S,2001A&A...373..916L,1980tsp..book.....C}:
\begin{eqnarray} 
\frac{\partial^2 \mathbold{\xi} }{\partial t^2}+ 2\eta_N \frac{\partial \mathbold{\xi} }{\partial t} +
{\cal L} \mathbold{\xi} = \frac{1}{\rho}  {\cal F}_{\rm conv}  + {\cal F}_{\rm gw}
\label{eq:motion}
\end{eqnarray} 
for the displacement $\mathbold{\xi} ({\bf r},t)$ of a forced oscillation corresponding
to a mode $N$.  In this equation, all the terms homogeneous in $\mathbold{\xi} $
have been put on the left-hand side, and the fluctuating terms arising from stochastic
excitation by turbulent convection ${\cal F}_{\rm conv}$ or by  GW
perturbations ${\cal F}_{\rm gw}$ are on the right-hand side. 
$\omega_N$ corresponds to the frequency of the mode $N$ and
$\rho$ is the density of the star in equilibrium\footnote{$N\equiv nlm$, where $n$, $l$ and $m$ are the order, 
degree, and azimuthal order of the mode.  In particular ,$n$ is a positive integer that relates with the number 
of nodes of $\xi_{\rm r}(r)$. 
As usual for modes with fixed $l$, $n=0$ is called the $f$-mode and $n\ge 1$ are the $p_n$ modes.
See~\citet{1989nos..book.....U} for details. In the remainder of the paper, if not stated otherwise, 
$N\equiv n2m $, where $m$ can be any integer such that $|m|\le 2$.   
}.
Although to compute the excitation, damping and propagation of acoustic and gravity 
waves inside stars it is necessary to resolve the full set of hydrodynamic equations, 
in the Sun and identical stars, the acoustic modes of oscillation are well 
represented by the linearised pulsation dynamics as described by the wave equation (\ref{eq:motion}). 
This equation has been very successful for explaining the solar and stellar 
observational data~\citep{2005MNRAS.360..859C}.

The pulsation variations of the fluid caused by momentum and heat  
are included in the damping rate $\eta_N$ and the linear spatial differential 
operator ${\cal L}$~\citep{1989nos..book.....U}.
Moreover,  these quantities are chosen  in such a way that both the frequency 
$\omega_N$ and eigenfunctions $\mathbold{\xi}_N ({\bf r})$  of the homogeneous equation 
\begin{eqnarray}
{\cal L}\mathbold{\xi}_N =\omega^2_N\;\mathbold{\xi}_N 
\end{eqnarray} 
are real. The set of eigenfunctions $\mathbold{\xi}_{N}$ 
can be shown to be orthogonal and form a complete set~\citep[e.g.,][]{1977Ap&SS..48..123A}.
In particular $\mathbold{\xi}_{\rm N}$ has two eigenfunction components
$\xi_{\rm r,N}(r)$ and $\xi_{\rm h,N}(r)$, the radial and horizontal surface displacements.
  
As already stated, we include as a source of excitation those fluctuations arising from turbulent
convection  $ {\cal F}_{\rm conv}({\bf r},t)$ which 
have been widely reported in the literature~\citep[e.g.,]{1977ApJ...211..934G,1994ApJ...424..466G,2008A&A...478..163B}, 
and  $ {\cal F}_{\rm gw}({\bf r},t)$ is the driving force related to GW fluctuations
of the spacetime continuum\footnote{Einstein notation. The Greek and Latin indices 
describe the coordinates in the spacetime manifold ($0,1,2,3$) and spatial coordinates ($1,2,3$).} 
where  the star is located~\citep{1973grav.book.....M}. 
$ {\cal F}_{\rm gw}({\bf r},t)$ has the components 
\begin{eqnarray} 
[{\cal F}_{\rm gw}]_i=1/2\; \ddot{h}_{ij}x^j.
\label{eq:Fgw}
\end{eqnarray}
$x^{j}$ are the spatial coordinates of index $j$ and  $\ddot{h}_{ij}$ 
is the second time derivative of the tensor  $h_{ij}$. As usual, $h_{ij}$ is the spatial part of the tensor
$h_{\alpha\beta}$ that describes a small perturbation relatively  to a  flat spacetime  universe (Minkowski space).
Moreover, the $h_{ij}$ deviation from a flat spacetime is solely attributed to GWs,  for which the effects of curvature is neglected due to the mass of the star~\citep[e.g.,][]{2009fcgr.book.....S}. 
 
Adopting a standard procedure of normal analysis~\citep[e.g.,][]{1989nos..book.....U}, we choose 
to represent any perturbation described by Equation (\ref{eq:motion}) 
as a combination of the eigenfunctions such that 
$\mathbold{\xi}_N ({\bf r},t)= A (t) \mathbold{\xi}_N ({\bf r})\;e^{-i\omega_{\rm N} t}$,
where $A(t)$ is the instantaneous amplitude of the mode~\citep{2005MNRAS.360..859C,2008A&A...478..163B}.
In  $\mathbold{\xi}_N ({\bf r},t)$ we do not show the term related with 
 the contribution of the temporal phase variation in the argument of $e^{-i\omega_{\rm N} t}$,
as this quantity is negligible for the formation of standing acoustic waves~\citep{2005MNRAS.360..859C}. 
Equally, the complex conjugate is also not represented as this quantity is not
relevant for our analysis~\citep{2001A&A...370..136S}.
This approximation is valid for modes for which the energy  exchange between the stellar turbulent convection and the oscillations  occur in a time-scale that is much longer than the oscillation period, i.e.,  $\eta_N\ll \omega_N$ as is the case of acoustic modes. This result has been shown to be valid for current solar and stellar acoustic oscillations.
By substituting this form of $\mathbold{\xi}_N ({\bf r},t)$ 
into Equation (\ref{eq:motion}), multiplying both members by $\mathbold{\xi}_N^{*}$ 
(the complex conjugate of $\mathbold{\xi}_N$\footnote{If not stated otherwise, throughout the remainder of article  
$\mathbold{\xi}_N$ will always refer to $\mathbold{\xi}_N$({\bf r}).}),  integrating this equation
for the total mass of the star and keeping only the leading terms, 
the equation reduces 
to
\begin{eqnarray} 
\frac{d^2A}{dt^2} +2\eta_{N} \frac{dA}{dt} +\omega_N^2 A = 
{\cal S_{\rm conv } } (t) + \delta^{l}_{2}\; {\cal S_{\rm gw } } (t),
\label{eq:Amotion}
\end{eqnarray}  
where $\delta^{l}_{2}$ is the  Kronecker tensor. 
Wave motion is a complex process with many second order terms. 
 Fortunately, these are very small when comparing with the leading terms, ${\cal S_{\rm conv } } (t) $
or ${\cal S_{\rm gw } } (t)$. Accordingly, the amplitudes of acoustic oscillations 
correspond to the solution of a damping harmonic oscillator as described by the previous equation.   
A detailed account about the nature of the second order terms
neglected in this computation can be found in~\citet{2005MNRAS.360..859C}. 

${\cal S_{\rm conv }}$ and  ${\cal S_{\rm gw } }$ are respectively the
excitation source terms related to turbulent convection and GWs. ${\cal S_{\rm gw } }$
reads 
\begin{eqnarray}
{\cal S_{\rm gw } } (t) =\frac{1}{I}
\int_0^R {\cal F}_{\rm gw} \cdot \mathbold{\xi}_N^*\; \rho r^2 dr,
\label{eq:Sgw}
\end{eqnarray}
where $R$ is the radius of the star and $I$ is the mode inertia.
I is an arbitrary constant which we choose to be equal to the mode of inertia,
as is usually done in the theory of stellar oscillations~\citep[e.g.,][]{2010aste.book.....A}.
$I$ is given by
\begin{eqnarray} 
I= 4\pi \int_0^R  \mathbold{\xi}_N\;\cdot \mathbold{\xi}_N^*\; \rho r^2 dr.
\label{eq:Mnorm}
\end{eqnarray}
It is convenient to introduce $M_N$, the so-called modal mass; 
thus $M_N=I/\zeta$, where $\zeta \equiv \xi_{r,N}^2(R)+6\xi_{h,N}^2(R)$.

In the eventuality of such a star having been perturbed by a passing GW, the response will be somehow 
identical to a tidal perturbation produced by a nearby object on the stellar  modes.  
Following from the specific properties of gravitational systems as  demonstrated in 
general relativity~\citep{Maggiore:2008tka}, GW perturbations only have modes with  $l\ge 2$. 
For convenience, we opt to study the leading order of the GW perturbation, i.e., 
the quadrupole modes ($l=2$). 
This is the reason why we have introduced $\delta^{l}_{2}$ in Equation (\ref{eq:Amotion}). 
\begin{figure}
\centering
\includegraphics[scale=0.5]{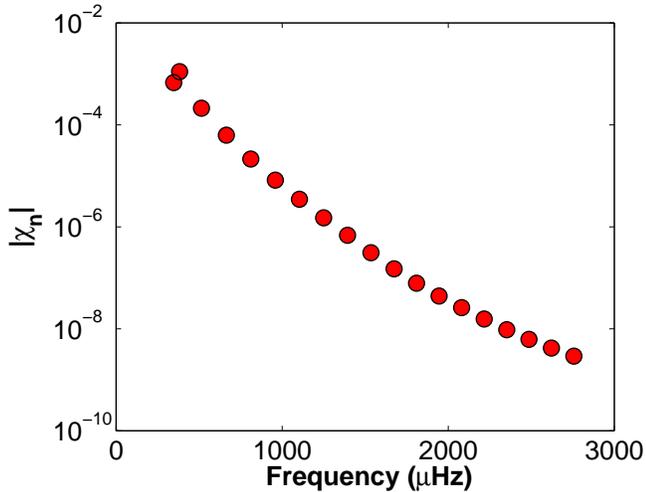}
\caption{Modulus $\chi_{n}$ coefficients for 
acoustic  quadrupole ($l=2$) modes of radial order $n$, from $0$ up to $18$.
The values of the  $\chi_{n}$ coefficients were computed for the current SSM.
The numerical values are shown in Table~\ref{tab:1}.
}
\label{fig:1}
\end{figure}

\begin{table} 
\centering
\caption{Quadrupole Acoustic Modes $(l=2)$ \\ Observational Data and  Standard Solar Model}
\begin{tabular}{llllllll}
\hline
\hline
n
&  Freq. [obs]~\footnote{The observational frequency table is obtained from  a compilation made by~\citet{2012RAA....12.1107T}, 
after the observations of ~\citet{2000ApJ...537L.143B,2001SoPh..200..361G,2004ApJ...604..455T,2009ApJS..184..288J}.
The strain $h_{-20}$ take values of $1$ to $10^{3}$.}    
& Freq.  [th]   &   $\chi_{n}$ &    $|L_n|$ & $V_{s,gw}$       \\
        &  ($\mu Hz$)           &   ($\mu Hz$)&   $\qquad$ & $({\rm cm })$ & $({\rm cm\; s^{-1}})$. \\
\hline
             &                        &      &  $\times 10^{-4}$& $\times 10^{7}$&  $\times h_{-20} 10^{-6}$ \\
$f$          &  $-$                   & $347.10 $  &  $-6.7432$ & $2.347$  &   $0.1884$ \\
 $p_1$       &  $-$                   & $382.26 $  &  $-11.038$ & $3.841$  &   $0.2673$ \\
 $p_2$       &  $-$                   & $514.48$   &  $+2.1193$ & $0.737$  &  $0.0169$ \\
 $p_3$       &  $-$                   & $664.06 $  &  $-0.6286$ & $0.219$  &  $0.0018$ \\
 $p_4$       &  $-$                   & $811.33 $  &  $+0.2133$ & $0.074$  &  $0.0003$ \\
\hline 
             &                        &           &  $\times 10^{-6}$& $\times 10^{5}$   &   $\times h_{-20} 10^{-10}$ \\
 $p_5$       &  $-$                   & $959.23 $  &  $-8.2377$ & $2.867$ &   $0.4484$ \\
 $p_6$       &  $-$                   & $1104.28$  &  $+3.4804$ & $1.211$ &   $0.0932$ \\
 $p_7$       &  $-$                   & $1249.78 $ &  $-1.5051$ & $0.524$ &   $0.0201$ \\
 $p_8$       & $1394.68\pm 0.01$      & $1393.68 $ &  $+0.6836$ & $0.238$ &   $0.0045$ \\
 $p_9$       & $1535.865\pm 0.006$    & $1535.08 $ &  $-0.3109$ & $0.108$ &   $0.0008$ \\
\hline 
             &                        &           &  $\times 10^{-8}$  &   $\times 10^4$ & $\times h_{-20} 10^{-15}$ \\ 
 $p_{10}$    & $1674.534\pm 0.013$    & $1673.80 $ &  $+14.946$  &         $1.082$ &   $22.670$\\
 $p_{11}$    & $1810.349\pm 0.015$    & $1809.40$   &  $-7.8242$       &   $0.520$ &   $7.7200$\\
 $p_{12}$    & $1945.800\pm 0.02$     & $1944.90$   &  $+4.3862 $       &   $0.272$ &  $3.1720$\\
 $p_{13}$    & $2082.150\pm 0.02$     & $2081.10 $   &  $-2.5981$       &   $0.153$&   $0.1413$ \\
 $p_{14}$    & $2217.69\pm 0.03$    & $2217.00 $       &  $+1.5564$       & $0.054$&   $0.7951$ \\
 $p_{15}$    & $2352.29\pm 0.03$    & $2352.30 $      &  $-0.9562$       &   $0.033$&  $0.4891$ \\
 $p_{16}$    & $2485.86\pm 0.03$    & $2486.60 $       &  $+0.6204$       &   $0.022$& $0.3085$ \\
 $p_{17}$    & $2619.64\pm 0.04$    & $2621.20 $       &  $-0.4180$       &   $0.014$& $0.1851$ \\
 $p_{18}$    & $2754.39\pm 0.04$    & $2756.90 $       &  $+0.2908$       &   $0.010$& $0.1275$ \\
\hline
\hline
\end{tabular}
\label{tab:1}
\end{table} 
 
Equation (\ref{eq:Sgw}) can be written in a more convenient form
by using  Equations (\ref{eq:Fgw}) and (\ref{eq:Mnorm}) for which 
${\cal S_{\rm gw } } (t)$ reads
\begin{eqnarray}
{\cal S_{\rm gw } } (t) = L_{n}\; \ddot{h}_{m}(t)
\label{eq:Sgwt}
\end{eqnarray}
where $L_{n}$ is the effective length that measures 
the sensitivity of a  mode of order $n$ to a GW perturbation 
and  $h_{\rm m} $ are the spherical components of $h_{ij}$ for which the
$m$ (azimuthal order) take one of the following integer values: $-2,-1,0,1,2$.
$L_{n}$ is given by  
\begin{eqnarray}
L_{n}=1/2\; R\; \chi_{n}
\label{eq:Lchi}
\end{eqnarray}
where $R$ is the radius of the star and $\chi_{n}$ is the coefficient that determines the
efficiency of a   mode of order $n$ to be excited by GWs. $\chi_{n}$ reads
\begin{eqnarray}
\chi_{n}=\frac{3}{4\pi \bar{\rho}_{\star}}
\int_0^1 \rho (r) \left[ \xi_{r,n2}(z)+3 \xi_{h,n2}(z) \right] r^3 dr.
\label{eq:chin}
\end{eqnarray}
In the computation of Equation (\ref{eq:Sgwt}), as is usually done, 
we arbitrarily normalized the eigenfunctions 
to the average density of the star $\bar{\rho}_{\star}$,  
such that $I\equiv (4\pi/3) R^3 \bar{\rho}_{\star}$. 
In the case of the Sun, $\bar{\rho}_{\star}$ is approximately $ 1.4\;{\rm g\;cm^{-3}}$.
Thus, Equation~(\ref{eq:chin}) is identical to others found in the literature, 
as by~\citet{1984ApJ...286..387B} and more recently by~\citet{2011ApJ...729..137S},
$\chi_{n}$ differ among these works only by the arbitrary normalization condition. 
Nevertheless, this theoretical model is developed in a similar manner to the one used for
resonant mass detectors. Thus, Equation (\ref{eq:Amotion}), in which the ${\cal S_{\rm conv } } (t)$ 
is neglected  and  $\bar{\rho}_{\star}$ is considered constant,  becomes equivalent 
to the one  found for a spherical resonant-mass detector~\citep[e.g.,][]{Maggiore:2008tka}.
This is the motivation for us to choose a normalization for 
$\chi_{n}$ that is identical to the one done for GW resonant-mass detectors.
 
Figure~\ref{fig:1} and Table~\ref{tab:1} show the $\chi_{n}$ coefficients computed for the
standard solar model~\citep[SSM:][]{1993ApJ...408..347T} with a stellar 
structure in very good agreement with helioseismology data.  
The difference between theoretical and observational  frequencies is smaller than 0.1\% (cf. Table~\ref{tab:1}).
This solar model was computed using a modified version of the {\sc Cesam} code~\citep{1997A&AS..124..597M}
for which the microphysics was updated.  In particular, we have computed the so-called 
low-Z SSM~\citep{2013ARA&A..51...21H} for which the solar composition used corresponds  
to the one determined by~\citet{2009ARA&A..47..481A}. 
The {\sc Cesam} nuclear physics network uses the fusion cross-sections recommended 
for the Sun by~\citet{2011RvMP...83..195A} with the most recent coefficients.
A detailed discussion about the physics of the current SSM can be found in the recent literature~\citep[e.g.,][]{2013MNRAS.435.2109L}. 

The values of $|\chi_{n}|$ in the Sun decrease with n (cf. Table~\ref{tab:1}),
a behavior identical to the one 
found for a resonance sphere of constant density.\footnote{Note that in the case of  
a sphere of constant density, $\chi_{n}$ depends only on the geometry of the star by means of the eigenfunctions
(cf. Equation (\ref{eq:chin})).}   However,  in the solar case,  $\chi_{n}$ is two orders of magnitude smaller. 
This difference is related to the fact that the solar density decreases
rapidly toward the Sun's surface and eigenfunctions of acoustic modes are more sensitive to the 
external layers of the star.  For instance, the largest of the $\chi_{n}$ coefficients,  
$\chi_{1}$ has a value of $-0.0011$ for the Sun  and $-0.328$ in the case of a resonant sphere~\citep{Maggiore:2008tka}. 
Moreover, $|L_{n}|$ takes values from $10^{7}\;{\rm cm }$  ($n=0$) to $100 \;{\rm cm }$ ($n=18$).
Solar low order modes have much larger values than the equivalent ones found in an  experimental detector. 
A similar quantity to $\chi_n$ was computed by~\citet{1984ApJ...286..387B} 
and by~\citet{2011ApJ...729..137S}. Unfortunately the comparison of $\chi_n$ for these models 
or a resonant-mass detector of constant mass as described by~\citet{Maggiore:2008tka} 
is not trivial to make. Nevertheless, $\chi_n$ varies in similar way to the $\chi_n$ factor 
found by~\citet{2011ApJ...729..137S}, in both cases these terms 
decrease as n increases and by identical orders of magnitude.

\begin{figure}
\centering
\includegraphics[scale=0.5]{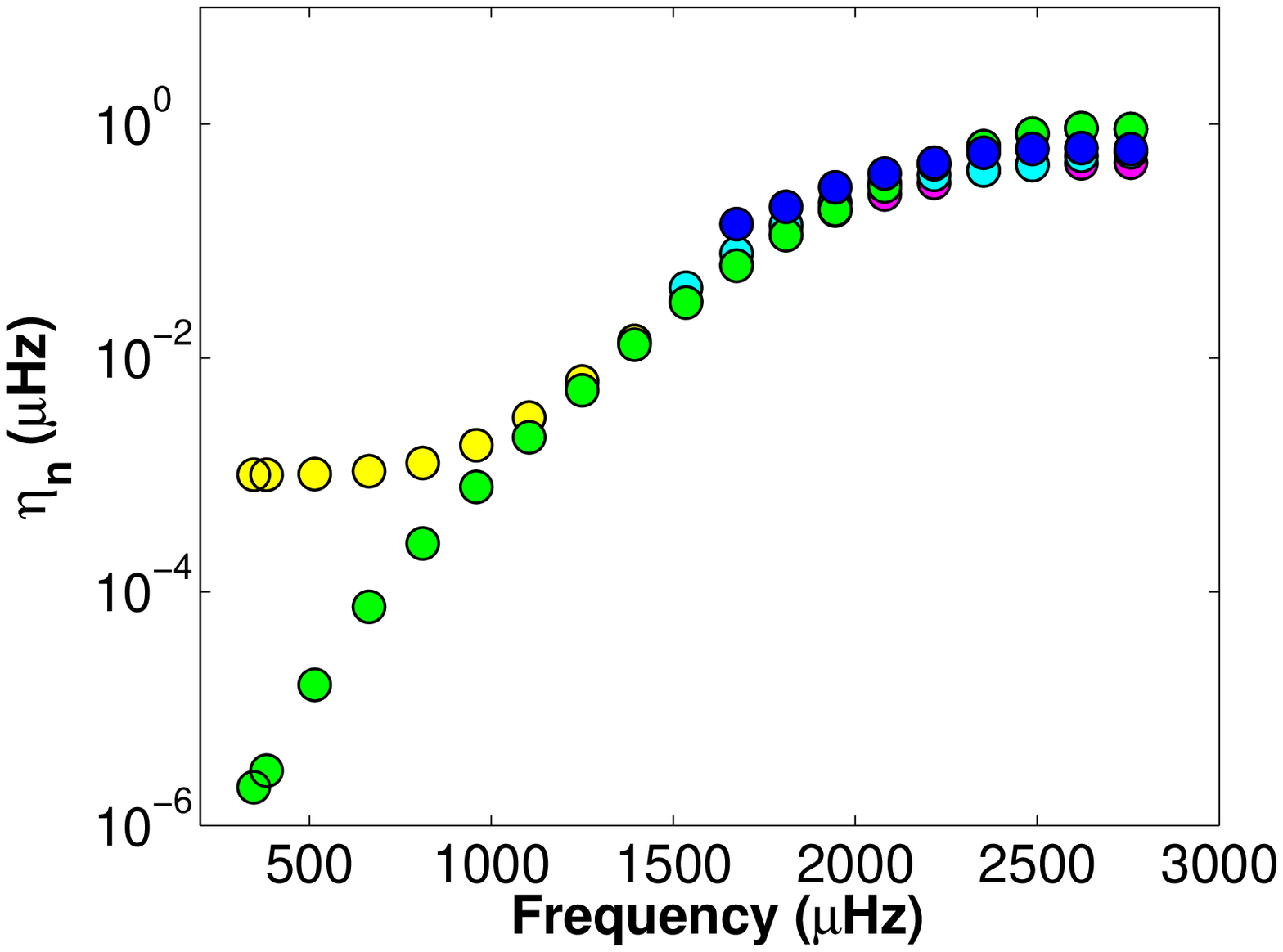}
\caption{Damping rates as a function of the frequency for the Sun.
The magenta, cyan and blue dots corresponds to the measurements made by~\citet{2005A&A...433..349B},~\citet{1997MNRAS.288..623C},
and~\citet{1988ApJ...334..510L}, and the green and yellow dots correspond to the theoretical predictions~\citep{1999A&A...351..582H,2002MNRAS.336L..65H,2009A&A...494..191B}.
The yellow dots corresponds to a "comparison" theoretical model
for which the damping rate is considered constant for $\nu \le 1.0 {\rm mHz}$.
The agreement between the theory and observation is  very good for the high frequencies, 
but for the lower frequencies no  observational data is available, 
and there are  only a few theoretical  predications. 
The  green and yellow  dots correspond to the values adopted for calculation of the 
GW transfer function (cf. Figure~\ref{fig:3}).}
\label{fig:2}
\end{figure}

\begin{figure}
\centering
\includegraphics[scale=0.5]{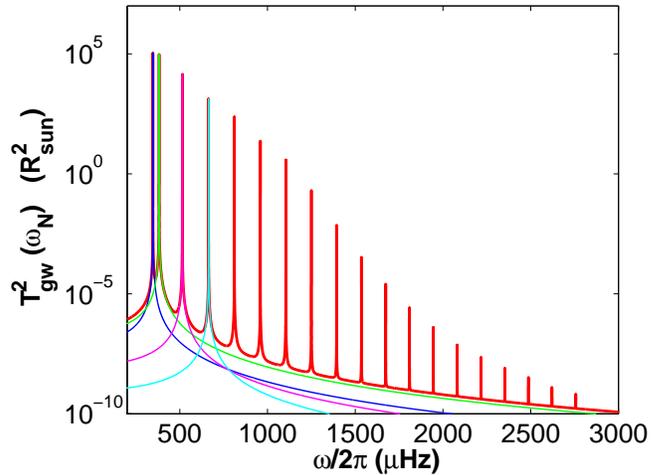}
\caption{
Square of the transfer function $T_{gw}^2 (\omega_N)$ for the acoustic quadrupole
modes of different radial order. All the acoustic modes show a clear well-defined
Lorentz profile. However, the low order modes have a larger
FWHM (full width at half-maximum) than the high order  modes.  
The red curve corresponds to the square transfer function of the combined 
quadrupole acoustic modes spectrum (yellow dots in Figure~\ref{fig:2}). 
The blue, green, magenta and cyan curves correspond
to $T_{gw}^2 (\omega_N)$ for  acoustic quadrupole modes of order $n=0,1,2$ and $3$.}
\label{fig:3}
\end{figure}

\section{Excitation of stellar modes by gravitational waves}
By taking the Fourier transform of Equation (\ref{eq:Amotion}) and neglecting
transient terms arising from the initial conditions on $A$, we obtain
for the averaged power spectrum $P_{N}(=\langle |\tilde{A}^2| \rangle)$:
\begin{eqnarray} 
P_{N}(\omega)=\frac{ P_{\rm conv} (\omega)+ \delta^{l}_{2}\; P_{\rm gw} (\omega) }{(\omega^2-\omega_{N}^2)^2+4\eta_{N}^2\omega^2}
\label{eq:Ptot}
\end{eqnarray}
where $P_{\rm conv}({\omega })= \langle |\tilde{S}_{\rm conv}^2| \rangle $  
and $P_{\rm gw}({\omega })= \langle |\tilde{S}_{\rm gw}^2| \rangle $
are the average power spectrum due to forcing caused by turbulent convection
and gravitational waves. $\tilde{f}(\omega)$ denotes the Fourier transform of $f(t)$.
This previous result is obtained under the approximation that 
the damping rate is always much smaller than the frequency, i.e., $|\eta_N|\ll \omega $,
as it is the case with acoustic oscillations of the Sun and Sun-like stars. 
In the derivation of the previous result, $P_{\rm conv} (\omega)$  and  $P_{\rm gw} (\omega)$  
are assumed to vary slowly with $\omega_{N}$. 

The power spectrum generated by stochastic excitation $P_{\rm conv} (\omega)$ 
is known to be caused by turbulent convection in the upper layers of the Sun and 
Sun-like stars just beneath the stellar photosphere~\citep[e.g.,][]{2008A&A...478..163B}.
This term represents the random spectrum due to the turbulent convection:  
if the temporal series is very long, the Lorentzian profile of each acoustic mode 
becomes visible due to the systematic beating of the mode 
by a random process of excitation~\citep{1995ESASP.376a.165K}.      
In the following, we compute  the GW contribution
to the power spectrum, i.e., $P_{N,{\rm gw} } (\omega)$.  
From Equations (\ref{eq:Sgwt}) and (\ref{eq:Ptot}), $P_{N,{\rm gw}} (\omega)$ reads
\begin{eqnarray} 
P_{N,{\rm  gw}}(\omega) = T^2_{N,{\rm  gw}}(\omega)\;P_{m}(\omega)
\label{eq:Pgw}
\end{eqnarray}
where $T_{N,{\rm  gw}}(\omega)$ is the transfer function of mode $N$
and $P_{m}(\omega)$ the power spectrum of the GW source. 
The former depends uniquely on the properties of the star, 
and the latter on the source of GWs. $T^2_{N,{\rm  gw}}(\omega)$ reads
\begin{eqnarray} 
T^2_{N,{\rm  gw}}(\omega) = 
\frac{L_{n}^2\; \omega^4}{(\omega^2-\omega_{N}^2)^2+4\eta_{N}^2\omega^2}.
\label{eq:Tgw}
\end{eqnarray}
The power spectrum of the GW source $P_{m}(\omega)$ is computed as $P_{m}(\omega)=|\tilde{h}_{m}|^2$.
In the Sun, the propagation of forward and backward traveling waves 
originating in the internal differential rotation 
leads to the generation of acoustic modes of different $m$.  
The frequency of these $m$-modes (fix $l$ and $n$) differs only by a few $\mu Hz$~\citep{2009LRSP....6....1H}. 
The solar magnetic field produces a similar effect leading 
to frequency differences of tens of $nHz$~\citep{2002ESASP.505...71A}.
Thus, for convenience, we will consider that $h_m$ and $P_{m}$
are fiducial values(for $l=2$ and $n$ fixed).  
This approximation is well justified as the different $h_m$ values 
mainly give us information about the direction of the GW source in 
the sky in relation to the star~\citep{Maggiore:2008tka}. 

In the following, we compute the rms surface velocity $V_N(\omega_N)$ of the $N$ mode, 
which is measured at a specific layer of the surface of the star~\citep[e.g.,][]{2001A&A...370..147S,2005MNRAS.360..859C}. 
Thus, the energy absorbed by a mode with a velocity $\dot{\xi}_N$ subject to
a force $F_{gw}(t)= M_N L_n \ddot{h}_m $ (Equation (\ref{eq:Amotion})), averaged  over several cycles, reads
\begin{eqnarray}
\frac{dE_{abs}}{dt} \equiv \langle F_{gw}(t) \dot{\xi}_N  \rangle=M_N h_o^2 \omega^2 \eta_N T^2_{N,{\rm gw}}(\omega). 
\label{eq:Eabs}
\end{eqnarray}
In this calculation, we consider that the gravitational wave source is monochromatic, $h_m=h_o\Re [e^{-i\omega t}]$,
where $h_o=10^{-17}h_{-17}$ is the strain sensitivity amplitude. In an experimental detector, $h_o$ is computed
from the strain spectral amplitude $h_f=h_o\sqrt{T}$, where $T$ is the observation time 
for a GW source that evolves slowly with time (source approximately monochromatic), 
or the characteristic width in the case of a short-lived GW burst.
In the case where $\omega\sim \omega_N$, Equation (\ref{eq:Eabs}) 
approaches the result ${dE_{abs}}/{dt} =2\eta_N E$, where $E$ is the energy of the mode. 
Therefore, the square of the surface rms velocity,  
$V_{N}^2 (\omega)\equiv {\zeta}/({2\eta_N I}) \;dE_{abs}/dt$\footnote{In the particular case of 
$V_{N}$ to be evaluated at $\omega=\omega_N$, this definition is equivalent to the one found 
in the literature~\citep[e.g.,][]{2001A&A...370..147S,2005MNRAS.360..859C}.} 
when excited by a GW source, reads 
\begin{eqnarray}
V_{N,gw}^2(\omega) = 
\frac{1/2\;\gamma_{\rm s}\; h_o^2\;L_{n}^2\; \omega^6}{(\omega^2-\omega_{N}^2)^2+4\eta_{N}^2\omega^2}
\label{eq:VN}
\end{eqnarray}
where $\gamma_{\rm s}$ is an additional parameter (dimensionless and of the order of unity),
which relates to the surface layer where the velocity measurement  is made. 
 
The oscillation quantities, such as the acoustic eigenfunctions,
strongly depend on the solar surface structure, especially the stellar atmosphere. 
Hence, to test the quality of our solar oscillation model, we  computed the normalized inertia 
${\cal E}_{nl} $ (with $l=2$) for the quadrupole acoustics modes,  
which are very sensitive to the surface of the star. We found that  
${\cal E}_{n2} $ varies from $5.8\times 10^{-4}$ for $n=0$ to $1.0\times 10^{-9}$ for $n=18$, 
these values are consistent with the results found in the literature~\citep{2000A&A...353..775P}.
In the case that $\omega=\omega_N$, Equation (\ref{eq:VN}) reduces to 
\begin{eqnarray}
V_{N,gw}^2(\omega_N)=\gamma_{\rm s}\frac{h_o^2 R^2\chi_n^2\omega_N^4}{32\eta_N^2}.
\label{eq:VNgwf}
\end{eqnarray}

\section{Discussion}

In the Sun, as in any spherical resonant-mass detector, the excitation of eigenmodes by an external GW source  
strongly depends of the internal structure of the star, and in particular on how these modes are damping 
in the stellar upper layers.
As shown in Equation (\ref{eq:VN}), $\eta_N$ is the leading coefficient that  
determines the capacity of solar acoustic oscillations to absorb GWs.
Although $\eta_N$ is determined with precision from solar oscillations    
in the high frequency range of the acoustic spectrum (above 1.5 ${\rm mHz}$),  
this is not the case in the lower frequency range. 
In this region of the spectrum, we only have a few theoretical predictions.

Figure~\ref{fig:2} shows the damping rates obtained by different observational groups:~\citet{1988ApJ...334..510L,1997MNRAS.288..623C,2005A&A...433..349B,2011JPhCS.271a2049G}, 
as well as the theoretical predictions of~\citet{1999A&A...351..582H,2005A&A...434.1055G,2009A&A...494..191B,2012A&A...540L...7B,2013ASPC..479...61B}.
The damping rate  increases in a nonlinear way with the frequency of the modes,
mostly due to the fact that  $\eta_N$  is strongly dependent on the properties of the convection
and the microphysics of the upper layers of the star~\citep[e.g.][]{2001MNRAS.322..473L,2014ApJ...782...16B}. 
The current predictions of $\eta_N$  agree well with observations for modes with $\nu \ge 1.5 {\rm  mHz}$.
Unfortunately, for modes in  the lower frequency range, observational data is non-existent, and
there are only a few theoretical predictions~\citep{1999A&A...351..582H,2009A&A...494..191B}. 
Estimating of the damping rate for low frequencies is very difficult. 

We note that Equation (\ref{eq:Amotion}) that describes the amplitude of acoustic oscillations  
 was obtained from the wave Equation (\ref{eq:motion}), which is a good
approximation for most of the acoustic oscillations~\citep{2005MNRAS.360..859C}. 
Moreover, even for such low values of $\eta_N$ the steady state solution is reached, 
even if it is not strictly the case for a pure harmonic damped oscillator~\citep[e.g.,][]{2004AAS...205.7605R,2014arXiv1405.1414M}. 
Actually, an $\eta_N$ of $\sim 10^{-2} \mu{\rm Hz}$ is currently observed for   
global low degree modes~\citep{1997MNRAS.288..623C,2005A&A...433..349B}. 
In particular, in the case of the Sun, the damping rates 
of all radial low order ($n\ge 1$ or $\nu\ge 250 \mu {\rm Hz}$)
have been successfully measured~\citep[e.g.,][]{2012RAA....12.1107T}. 
As acoustic modes with  similar frequencies are equally damped in the convection zone, 
the damping rates of quadrupole modes can be estimated from the same quantities 
measured from radial modes.

This is due to the fact that  hydrodynamic simulations of turbulent convection in stars 
are not able to accurately reproduce stellar convection. As a consequence, the prediction 
of damping and excitation of low order modes, including the damping of quadrupole acoustic modes, 
is not fully reliable. For future use, in Figure~\ref{fig:2} we show a "comparison" model  
in which  $\eta_N$ is almost constant for $\nu \le 1 {\rm mHz}$, and the damping rate
of low order modes is assumed to be identical to the $\eta_N$ value for $\nu \sim 1 {\rm mHz}$.
The motivation for representing this "comparison" theoretical model  
is to show the importance of $\eta_N$ in the detection of GW events. 
In particular, the value of $\eta_N$  for low values of $\nu$ 
has a major impact on the transfer function.

Figure~\ref{fig:3} shows $T^2_{N,{\rm gw}}(\omega)$ for the acoustic quadrupole modes.
The quality of  $T^2_{N,{\rm gw}}(\omega)$ for each mode 
can be measured by the quality factor $Q_N\equiv\omega_N/(2\eta_N)$. 
In the Sun, $Q_N$ varies from $10^8$ to $10^4$ running from $n=0$ until $n=18$.
In particular, the  $Q_N$ of low order modes is higher than the value found for
the most advanced resonant spherical detectors~\citep[$\sim 10^{6}-10^{7} $,][]{2007PhRvD..75b2002G}.
This high quality factor is the reason why the Lorentz profile is almost $\delta$-function-like,
as needed for an ideal GW detector (cf. Figure~\ref{fig:3}). 
$T^2_{N,{\rm gw}}(\omega)$ also depends on the value of $\chi_n$ (cf. Table~\ref{tab:1}). 
$T^2_{N,{\rm gw}}(\omega)$ decreases with increasing $n$. 
This is also found in the case of a sphere with constant density,  nevertheless, 
the variation of $T^2_{N,{\rm gw}}(\omega)$ with $n$ is more pronounced
due to the fact that unlike in a detector, the density inside the Sun is not constant.   
 
\begin{figure}
\centering
\includegraphics[scale=0.5]{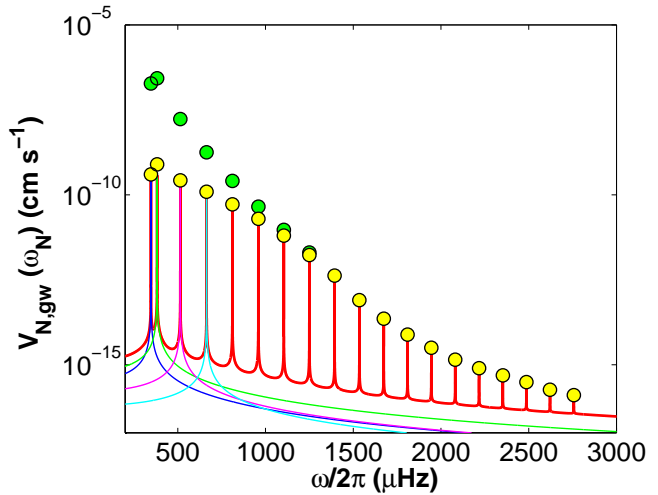}
\caption{
Velocity power spectrum of the quadrupole modes of different orders exited by
an external GW source excited by a fiducial strain   of $h_{-20}=1$: 
the peaks occurs at the location of eigenfrequencies
$\nu_{nl}$ corresponding to the different acoustic eigenmodes 
of the Sun (cf. Table~\ref{tab:1}). The red curve corresponds to the combined
power spectrum (Equation (\ref{eq:VN})). The  blue, green, magenta and cyan curves correspond
to the power spectrum of the  acoustic eigenmodes of order $n=0,1,2$ and $3$.
The green and yellow dots (Equation (\ref{eq:VNgwf})) correspond to the two sets of theoretical $\eta_N$ 
values shown in Figure~\ref{fig:3}.}
\label{fig:4}
\end{figure}

In the Sun,  acoustic oscillations of low order are driven by stochastic turbulent convection,  
which leads to a well-defined value of the rms velocity at the solar surface for each mode. 
In the case with quadrupole modes, as shown in the previous section, the  
rms velocity also has an additional GW component as predicted by Equation (\ref{eq:VN}).
Figure~\ref{fig:4} shows the predicted $V_{N,{\rm gw}}(\omega)$, assumed to be 
excited by a GW source with a fiducial strain amplitude of $h_{-20}=1$. 
Equation (\ref{eq:VN}) defines the profile of the GW-excited mode profile, 
and Equation (\ref{eq:VNgwf}) defines the value of $V_N^2{\omega}$  at the acoustic frequency mode. 
$V_{N,{\rm gw}}(\omega)$ shows a $\delta$-function-like profile 
as already found in the $T^2_{N,{\rm gw}}(\omega)$ (cf. Figure~\ref{fig:4}).

\citet{2011ApJ...729..137S} have obtained an expression for the rms velocity
amplitude similar to Equation (\ref{eq:VNgwf}) using a different 
formulation for the excitation of quadrupole modes by  gravitational waves.   
Our predictions of $V_{N,{\rm gw}}(\omega)$ for the lower 
order acoustic modes ($n\le 5$) for which we consider that $\eta_N\sim 10^{-3}\;\mu Hz$
and  $h_{-20}=1$ are identical to the predictions of~\citet{2011ApJ...729..137S}.
Nevertheless, we notice that for the lower order acoustic modes
the theoretical predictions of damping rates decrease with decreasing n ($0 \le n\le 4$), 
from $10^{-3}$ to $10^{-6}\,\mu {\rm Hz}$ (Cf. Figure~\ref{fig:2}), 
for which  $V_{N,{\rm gw}}(\omega)$ varies from $10^{-9}$ to $10^{-6}\,{\rm cm\, s^{-1}}$
(Cf. Figure~\ref{fig:4}). 

If $\eta_N$ has values of the order of $10^{-6}\;\mu{\rm Hz}$ or $ 10^{-3}\;\mu {\rm Hz} $
as predicted by some theoretical damping oscillation models (cf. Figure~\ref{fig:2}), 
GW events with $h_{-20}$ lead to 
$V_{N,{\rm gw}}(\omega)$ with  $10^{-9}\;{\rm cm s^{-1}} $ (comparison model)  or 
$10^{-6}\;{\rm cm s^{-1}} $ (theoretical model).
In the case of an occurrence of  GW events with $h_{-20}\sim 10^3$, $V_{N,{\rm gw}}(\omega)$ will have values
of the order of $10^{-6}\;{\rm cm s^{-1}} $ or $10^{-3} \;{\rm cm s^{-1}} $.
This latter result is relatively near the current helioseismology measurements.

In principle, it should be possible to separate the quadrupole excitation by gravitational waves
from the excitation by convection. 
Current observational data of  helio- and asteroseismology allows us to determine in great detail 
 the properties of damping and excitation of acoustic oscillations by the turbulent motions in the stellar upper layers~\citep[e.g.,][]{2001MNRAS.322..473L}.
In particular, the accurate measurement of frequencies, damping rates and the maximum rms surface velocities of global
acoustic modes (modes with $l\le 4$) can be used to separate the GW excitation of quadrupole modes from the excitation and damping due to the turbulent convection. This is possible because it  has been shown both theoretically and observationally
that the excitation and damping of global acoustic modes by convection  (including quadruple $l= 2$)
depends only on the frequency of the mode (and is independent of the degree of the mode).
As all the low degree modes are equally excited by convection, if a low order quadrupole is stimulated by a GW source, 
it will show an unique pattern in the pulsation spectrum,  quite distinct from the other global acoustic modes (like radial, dipole and octopoles) with identical frequencies. 
This should be a strong hint of excitation of quadrupole modes by a GW source.

\section{Summary and Conclusion}

In this article, we calculated the excitation of acoustic quadrupole modes by GW in a 
star like the Sun by using a formulation identical to that used for the computation of 
eigenmodes in resonant-mass detectors.
In this work, we have use realistic theoretical predictions
of damping rates for acoustic modes of low order which have been validated at high 
frequencies.

In particular, we find that the low-order modes in the Sun 
have a quality factor an order of magnitude higher than 
those found in resonant-mass detectors.   
Moreover, the sensitivity of acoustic modes to GW perturbations is regulated by an 
effective length as in an experimental bar/sphere detector which in the Sun takes values 
between $10^7$ cm and $10$ cm. 
This large variation in the value of the effective length is related to the fact that in stars, 
the eigenfunctions of acoustic modes (increasing with the order of the mode) are 
mostly sensitive to the stellar envelope and less sensitive to the stellar core. 

The helioseismological acoustic wave frequencies overlap with the gravitational radiation frequency range that will be probed by eLISA~\citep{2013GWN.....6....4A}. 
One of the targets of eLISA will be nearby ultracompact  binaries. The sensitivity $h_f$ of eLISA will be only 
$10^{-18}\rm (Hz)^{-1/2}$ at 0.001 Hz, and a factor of 10 worse at 0.0003 Hz.
The brightest nearby binaries have predicted strain spectral amplitudes in the range 
($3.10^{-18}-3.10^{-17})\rm (Hz)^{-1/2}$
over frequencies 0.01Hz to 0.001Hz.The strongest binaries over two years of observation are predicted to have 
$h_f \sim 10^{-17} \rm (Hz)^{-1/2}$ (or $h_o \sim 10^{-20}$)  and frequencies as low as 0.0003Hz.
The helioseismological modes are excited over  300-3000 $\mu$Hz and could be up to a 
factor 100 more sensitive than eLISA.\footnote{We remind the reader 
that $h_f=h_o/\sqrt{T}$ where $T$ in the is the observation time (see Section~3).}

Presently, the main caveat in this model is the damping rate, which in the case with 
modes with high frequencies is well determined ($\nu \ge 1.5  {\rm mHz}$) from observations,
but in the case with modes with low frequencies the damping rates are theoretical. 
Accordingly, with present damping  rate estimates, we predict an rms square velocity on the solar surface of 
the order of $(10^{-1}$-- $1)h_{-20}$ ${\rm cm\, s^{-1}}$ for an GW event with a strain amplitude of $10^{-20}h_{-20}$. 
Some of these values are near the current rms surface velocity amplitudes measured in the Sun's surface. 

In principle, as in experimental detectors, the measurement of the maximum
amplitude of rms velocity of quadrupole eigenmodes excited by  
GW periodic or random events is very difficult.
Nevertheless, this difficulty could be in part be  overcome by taking advantage of several 
aspects  that are unique to stars: (1) stars (due to their very large masses) 
have a very high GW integrated cross-section; 
(2) a large number of stars of different 
masses have been found (presently more than 500) to oscillate in a manner identical to the Sun; 
(3) stellar seismology instruments are recording very long 
time series of seismic data, in some cases spanning over several years, and  
in the case of the Sun more than two decades; (4) the possibility of looking 
simultaneously for the same single or periodic GW event in distinct 
stars (as GWs propagates between  stars at 
the speed of light); and (5) the possibility of using radial 
and dipole acoustic modes to isolate the GW signal in the 
quadrupole mode, as the excitation and damping of 
acoustic modes depends uniquely on the frequency.
In particular, oscillating stars can provide a unique way to look for contemporaneous quadrupole mode excitations
in different stars by a single GW event.  As the distances between many of  these stars are relatively small, as in the case of stellar clusters, this can be used advantageously to look for the same GW imprint on quadrupole modes of different stars. In these cases, the time-lag between the excitation of quadrupole modes of two distinct stars can be determined accurately from the locations of the stars and the speed of propagation of the GWs.

Although the challenges are great, the discovery of GW via stellar acoustic oscillations by the current set-up of experiments 
on  Earth and/or in space is such an exceptional outcome that all the effort toward 
accomplishing this goal is well worth the investment.


\begin{acknowledgments}
The authors would like to thanks comments and suggestions  of the anonymous referee, as well of 
the many colleagues that have commented and make suggestions to
a preliminary version of the manuscript. We are particular grateful to Coleman Miller, Barry McKernan, 
Zoltan Haiman, Daniel Siegel, Markus Roth and Bernard Schutz.
The authors thank J. Moore and colleagues for maintaining the website: 
Gravitational Wave Detectors and Sources (http://www.ast.cam.ac.uk/~rhc26/sources/).

The work of I.L. was supported by grants from "Funda\c c\~ao para a Ci\^encia e Tecnologia"  and "Funda\c c\~ao Calouste Gulbenkian". The research of J.S. has been supported at IAP by  ERC project  267117 (DARK) hosted by Universit\'e Pierre 
et Marie Curie -- Paris 6 and at JHU by NSF grant OIA-1124403.
We are grateful to the authors of ADIPLS and CESAM codes for having made their codes publicly available.
\end{acknowledgments}
\bibliographystyle{yahapj}

\begin{thebibliography}{64}
\expandafter\ifx\csname natexlab\endcsname\relax\def\natexlab#1{#1}\fi

\bibitem[{Abbott {et~al.}(2009)Abbott, Abbott, Adhikari, Ajith, Allen, Allen,
  Amin, Anderson, Anderson, Arain, Araya, Armandula, Armor, Aso, Aston,
  Aufmuth, Aulbert, Babak, Baker, Ballmer, Barker, Barker, Barr, Barriga,
  Barsotti, Barton, Bartos, Bassiri, Bastarrika, Behnke, Benacquista,
  Betzwieser, Beyersdorf, Bilenko, Billingsley, Biswas, Black, Blackburn,
  Blackburn, Blair, Bland, Bodiya, Bogue, Bork, Boschi, Bose, Brady, Braginsky,
  Brau, Bridges, Brinkmann, Brooks, Brown, Brummit, Brunet, Bullington,
  Buonanno, Burmeister, Byer, Cadonati, Camp, Cannizzo, Cannon, Cao, Cardenas,
  Caride, Castaldi, Caudill, Cavagli{\`a}, Cepeda, Chalermsongsak, Chalkley,
  Charlton, Chatterji, Chelkowski, Chen, Christensen, Chung, Clark, Clark,
  Clayton, Cokelaer, Colacino, Conte, Cook, Corbitt, Cornish, Coward, Coyne,
  Creighton, Creighton, Cruise, Culter, Cumming, Cunningham, Danilishin,
  Danzmann, Daudert, Davies, Daw, DeBra, Degallaix, Dergachev, Desai, DeSalvo,
  Dhurandhar, Diaz, Dietz, Donovan, Dooley, Doomes, Drever, Dueck, Duke, Dumas,
  Dwyer, Echols, Edgar, Effler, Ehrens, Espinoza, Etzel, Evans, Evans,
  Fairhurst, Faltas, Fan, Fazi, Fehrmenn, Finn, Flasch, Foley, Forrest,
  Fotopoulos, Franzen, Frede, Frei, Frei, Freise, Frey, Fricke, Fritschel,
  Frolov, Fyffe, Galdi, Garofoli, Gholami, Giaime, Giampanis, Giardina, Goda,
  Goetz, Goggin, Gonz{\'a}lez, Gorodetsky, Go{\ss}ler, Gouaty, Grant, Gras,
  Gray, Gray, Greenhalgh, Gretarsson, Grimaldi, Grosso, Grote, Grunewald,
  Guenther, Gustafson, Gustafson, Hage, Hallam, Hammer, Hammond, Hanna, Hanson,
  Harms, Harry, Harry, Harstad, Haughian, Hayama, Heefner, Heng, Heptonstall,
  Hewitson, Hild, Hirose, Hoak, Hodge, Holt, Hosken, Hough, Hoyland, Hughey,
  Huttner, Ingram, Isogai, Ito, Ivanov, Johnson, Johnson, Jones, Jones, Jones,
  Ju, Kalmus, Kalogera, Kandhasamy, Kanner, Kasprzyk, Katsavounidis, Kawabe,
  Kawamura, Kawazoe, Kells, Keppel, Khalaidovski, Khalili, Khan, Khazanov,
  King, Kissel, Klimenko, Kokeyama, Kondrashov, Kopparapu, Koranda, Kozak,
  Krishnan, Kumar, Kwee, Lam, Landry, Lantz, Lazzarini, Lei, Lei, Leindecker,
  Leonor, Li, Lin, Lindquist, Littenberg, Lockerbie, Lodhia, Longo, Lormand,
  Lu, Lubinski, Lucianetti, L{\"u}ck, Machenschalk, MacInnis, Mageswaran,
  Mailand, Mandel, Mandic, M{\'a}rka, M{\'a}rka, Markosyan, Markowitz, Maros,
  Martin, Martin, Marx, Mason, Matichard, Matone, Matzner, Mavalvala, McCarthy,
  McClelland, McGuire, McHugh, McIntyre, McKechan, McKenzie, Mehmet, Melatos,
  Melissinos, Men{\'e}ndez, Mendell, Mercer, Meshkov, Messenger, Meyer, Miller,
  Minelli, Mino, Mitrofanov, Mitselmakher, Mittleman, Miyakawa, Moe, Mohanty,
  Mohapatra, Moreno, Morioka, Mors, Mossavi, Mow~Lowry, Mueller,
  M{\"u}ller-Ebhardt, Muhammad, Mukherjee, Mukhopadhyay, Mullavey, Munch,
  Murray, Myers, Myers, Nash, Nelson, Newton, Nishizawa, Numata, O'Dell,
  O'Reilly, O'Shaughnessy, Ochsner, Ogin, Ottaway, Ottens, Overmier, Owen, Pan,
  Pankow, Papa, Parameshwaraiah, Patel, Pedraza, Penn, Perraca, Pierro, Pinto,
  Pitkin, Pletsch, Plissi, Postiglione, Principe, Prix, Prokhorov, Punken,
  Quetschke, Raab, Rabeling, Radkins, Raffai, Raics, Rainer, Rakhmanov,
  Raymond, Reed, Reed, Rehbein, Reid, Reitze, Riesen, Riles, Rivera, Roberts,
  Robertson, Robinson, Robinson, Roddy, R{\"o}ver, Rollins, Romano, Romie,
  Rowan, R{\"u}diger, Russell, Ryan, Sakata, de~la Jordana, Sandberg,
  Sannibale, Santamar{\'\i}a, Saraf, Sarin, Sathyaprakash, Sato, Satterthwaite,
  Saulson, Savage, Savov, Scanlan, Schilling, Schnabel, Schofield, Schulz,
  Schutz, Schwinberg, Scott, Scott, Searle, Sears, Seifert, Sellers, Sengupta,
  Sergeev, Shapiro, Shawhan, Shoemaker, Sibley, Siemens, Sigg, Sinha, Sintes,
  Slagmolen, Slutsky, Smith, Smith, Smith, Somiya, Sorazu, Stein, Stein,
  Steplewski, Stochino, Stone, Strain, Strigin, Stroeer, Stuver, Summerscales,
  Sun, Sung, Sutton, Szokoly, Talukder, Tang, Tanner, Tarabrin, Taylor, Taylor,
  Thacker, Thorne, Th{\"u}ring, Tokmakov, Torres, Torrie, Traylor, Trias,
  Ugolini, Ulmen, Urbanek, Vahlbruch, Vallisneri, Van Den~Broeck, van~der
  Sluys, van Veggel, Vass, Vaulin, Vecchio, Veitch, Veitch, Veltkamp, Villar,
  Vorvick, Vyachanin, Waldman, Wallace, Ward, Weidner, Weinert, Weinstein,
  Weiss, Wen, Wen, Wette, Whelan, Whitcomb, Whiting, Wilkinson, Willems,
  Williams, Williams, Willke, Wilmut, Winkelmann, Winkler, Wipf, Wiseman, Woan,
  Wooley, Worden, Wu, Yakushin, Yamamoto, Yan, Yoshida, Zanolin, Zhang, Zhang,
  Zhao, Zotov, Zucker, M{\"u}hlen, \& Zweizig}]{2009RPPh...72g6901A}
Abbott, B.~P., Abbott, R., Adhikari, R., {et~al.} 2009,
  \href{http://dx.doi.org/10.1088/0034-4885/72/7/076901}{Reports on Progress in
  Physics, 72, 6901}

\bibitem[{Adelberger {et~al.}(2011)Adelberger, Garc{\'\i}a, Robertson, Snover,
  Balantekin, Heeger, Ramsey-Musolf, Bemmerer, Junghans, Bertulani, Chen,
  Costantini, Prati, Couder, Uberseder, Wiescher, Cyburt, Davids, Freedman,
  Gai, Gazit, Gialanella, Imbriani, Greife, Hass, Haxton, Itahashi, Kubodera,
  Langanke, Leitner, Leitner, Vetter, Winslow, Marcucci, Motobayashi,
  Mukhamedzhanov, Tribble, Nollett, Nunes, Park, Parker, Schiavilla, Simpson,
  Spitaleri, Strieder, Trautvetter, Suemmerer, \& Typel}]{2011RvMP...83..195A}
Adelberger, E.~G., Garc{\'\i}a, A., Robertson, R. G.~H., {et~al.} 2011,
  \href{http://dx.doi.org/10.1103/RevModPhys.83.195}{Review of Modern Physics,
  83, 195}

\bibitem[{{Aerts} {et~al.}(2010){Aerts}, {Christensen-Dalsgaard}, \&
  {Kurtz}}]{2010aste.book.....A}
{Aerts}, C., {Christensen-Dalsgaard}, J., \& {Kurtz}, D.~W. 2010,
{Asteroseismology}, 
\href{http://adsabs.harvard.edu/abs/2010aste.book.....A}{
Astronomy and Astrophysics Library}. 
Springer Science+Business Media B.V.


\bibitem[{Aizenman \& Smeyers(1977)}]{1977Ap&SS..48..123A}
Aizenman, M.~L., \& Smeyers, P. 1977,
  \href{http://dx.doi.org/10.1007/BF00643044}{Astrophysics and Space Science,
  48, 123}

\bibitem[{{Amaro-Seoane} {et~al.}(2013){Amaro-Seoane}, {Aoudia}, {Babak},
  {Bin{\'e}truy}, {Berti}, {Boh{\'e}}, {Caprini}, {Colpi}, {Cornish},
  {Danzmann}, {Dufaux}, {Gair}, {Hinder}, {Jennrich}, {Jetzer}, {Klein},
  {Lang}, {Lobo}, {Littenberg}, {McWilliams}, {Nelemans}, {Petiteau}, {Porter},
  {Schutz}, {Sesana}, {Stebbins}, {Sumner}, {Vallisneri}, {Vitale},
  {Volonteri}, {Ward}, \& {Wardell}}]{2013GWN.....6....4A}
{Amaro-Seoane}, P., {Aoudia}, S., {Babak}, S., {et~al.} 2013, 
\href{http://adsabs.harvard.edu/abs/2013GWN.....6....4A}{GW Notes, Vol.~6, p.~4-110, 6, 4}

\bibitem[{Antia(2002)}]{2002ESASP.505...71A}
Antia, H.~M. 2002,Proceedings of the Magnetic Coupling of the Solar Atmosphere
  Euroconference and IAU Colloquium 188,  Ed. H. Sawaya-Lacoste. ESA SP-505. Noordwijk, Netherlands: ESA Publications Division, \href{http://adsabs.harvard.edu/abs/2002ESASP.505...71A}{p. 71 – 78, 71}

\bibitem[{Asplund {et~al.}(2009)Asplund, Grevesse, Sauval, \&
  Scott}]{2009ARA&A..47..481A}
Asplund, M., Grevesse, N., Sauval, A.~J., \& Scott, P. 2009,
  \href{http://dx.doi.org/10.1146/annurev.astro.46.060407.145222}{Annual Review
  of Astronomy and Astrophysics, 47, 481}

\bibitem[{Baudin {et~al.}(2005)Baudin, Samadi, Goupil, Appourchaux, Barban,
  Boumier, Chaplin, \& Gouttebroze}]{2005A&A...433..349B}
Baudin, F., Samadi, R., Goupil, M.~J., {et~al.} 2005,
  \href{http://dx.doi.org/10.1051/0004-6361:20041229}{Astronomy and
  Astrophysics, 433, 349}

\bibitem[{Belkacem {et~al.}(2012)Belkacem, Dupret, Baudin, Appourchaux,
  Marques, \& Samadi}]{2012A&A...540L...7B}
Belkacem, K., Dupret, M.~A., Baudin, F., {et~al.} 2012,
  \href{http://dx.doi.org/10.1051/0004-6361/201218890}{Astronomy and
  Astrophysics, 540, L7}

\bibitem[{Belkacem {et~al.}(2008)Belkacem, Samadi, Goupil, \&
  Dupret}]{2008A&A...478..163B}
Belkacem, K., Samadi, R., Goupil, M.~J., \& Dupret, M.~A. 2008,
  \href{http://dx.doi.org/10.1051/0004-6361:20077775}{Astronomy and
  Astrophysics, 478, 163}

\bibitem[{Belkacem {et~al.}(2009)Belkacem, Samadi, Goupil, Dupret, Brun, \&
  Baudin}]{2009A&A...494..191B}
Belkacem, K., Samadi, R., Goupil, M.~J., {et~al.} 2009,
  \href{http://dx.doi.org/10.1051/0004-6361:200810827}{Astronomy and
  Astrophysics, 494, 191}

\bibitem[{Belkacem {et~al.}(2013)Belkacem, Samadi, Mosser, Goupil, \&
 Ludwig}]{2013ASPC..479...61B}
Belkacem, K., Samadi, R., Mosser, B., Goupil, M.~J., \& Ludwig, H.~G. 2013,
in Progress in Physics of the Sun and Stars: A New Era in Helio- and
Asteroseismology.  Edited by H. Shibahashi and A.E. Lynas-Gray, San Francisco: Astronomical Society of the Pacific,
\href{http://adsabs.harvard.edu/abs/2013ASPC..479...61B}{ASP Conference Proceedings, Vol. 479, 61}
  
  
\bibitem[{Bertello {et~al.}(2000)Bertello, Varadi, Ulrich, Henney, Kosovichev,
  Garcia, \& Turck-Chieze}]{2000ApJ...537L.143B}
Bertello, L., Varadi, F., Ulrich, R.~K., {et~al.} 2000,
  \href{http://dx.doi.org/10.1086/312775}{The Astrophysical Journal, 537, L143}

\bibitem[{Boughn \& Kuhn(1984)}]{1984ApJ...286..387B}
Boughn, S.~P., \& Kuhn, J.~R. 1984,
  \href{http://dx.doi.org/10.1086/162612}{Astrophysical Journal, 286, 387}

\bibitem[{Brito \& Lopes(2014)}]{2014ApJ...782...16B}
Brito, A., \& Lopes, I. 2014,
  \href{http://dx.doi.org/10.1088/0004-637X/782/1/16}{The Astrophysical
  Journal, 782, 16}

\bibitem[{Chaplin {et~al.}(1997)Chaplin, Elsworth, Isaak, McLeod, Miller, \&
  New}]{1997MNRAS.288..623C}
Chaplin, W.~J., Elsworth, Y., Isaak, G.~R., {et~al.} 1997,
  \href{http://adsabs.harvard.edu/abs/1997MNRAS.288..623C}{Monthly Notices of
  the Royal Astronomical Society, 288, 623}

\bibitem[{Chaplin {et~al.}(2005)Chaplin, Houdek, Elsworth, Gough, Isaak, \&
  New}]{2005MNRAS.360..859C}
Chaplin, W.~J., Houdek, G., Elsworth, Y., {et~al.} 2005,
  \href{http://dx.doi.org/10.1111/j.1365-2966.2005.09041.x}{Monthly Notices of
  the Royal Astronomical Society, 360, 859}

\bibitem[{Chaplin \& Miglio(2013)}]{2013ARA&A..51..353C}
Chaplin, W.~J., \& Miglio, A. 2013,
  \href{http://dx.doi.org/10.1146/annurev-astro-082812-140938}{Annual Review of
  Astronomy and Astrophysics, 51, 353}

\bibitem[{Chaplin {et~al.}(2011)Chaplin, Kjeldsen, Christensen-Dalsgaard, Basu,
  Miglio, Appourchaux, Bedding, Elsworth, Garc{\'\i}a, Gilliland, Girardi,
  Houdek, Karoff, Kawaler, Metcalfe, Molenda-{\.{Z}}akowicz, Monteiro,
  Thompson, Verner, Ballot, Bonanno, Brand{\~a}o, Broomhall, Bruntt, Campante,
  Corsaro, Creevey, Dogan, Esch, Gai, Gaulme, Hale, Handberg, Hekker, Huber,
  Jim{\'e}nez, Mathur, Mazumdar, Mosser, New, Pinsonneault, Pricopi, Quirion,
  R{\'e}gulo, Salabert, Serenelli, Silva~Aguirre, Sousa, Stello, Stevens,
  Suran, Uytterhoeven, White, Borucki, Brown, Jenkins, Kinemuchi, Van~Cleve, \&
  Klaus}]{2011Sci...332..213C}
Chaplin, W.~J., Kjeldsen, H., Christensen-Dalsgaard, J., {et~al.} 2011,
  \href{http://dx.doi.org/10.1126/science.1201827}{Science, 332, 213}

\bibitem[{Cox(1980)}]{1980tsp..book.....C}
Cox, J.~P. 1980, {Theory of stellar pulsation} (Research supported by the
  National Science Foundation Princeton)

\bibitem[{Dyson(1969)}]{1969ApJ...156..529D}
Dyson, F.~J. 1969, \href{http://dx.doi.org/10.1086/149986}{Astrophysical
  Journal, 156, 529}

\bibitem[{Gair \& Porter(2012)}]{2012arXiv1210.8066G}
Gair, J.~R., \& Porter, E.~K. 2012,
  \href{http://arxiv.org/abs/1210.8066}{eprint arXiv:1210.8066}

\bibitem[{Gair {et~al.}(2013)Gair, Vallisneri, Larson, \&
  Baker}]{2013LRR....16....7G}
Gair, J.~R., Vallisneri, M., Larson, S.~L., \& Baker, J.~G. 2013,
  \href{http://dx.doi.org/10.12942/lrr-2013-7}{Living Reviews in Relativity,
  16, 7}

\bibitem[{Garcia {et~al.}(2011)Garcia, Salabert, Ballot, Sato, Mathur, \&
  Jimenez}]{2011JPhCS.271a2049G}
Garcia, R.~A., Salabert, D., Ballot, J., {et~al.} 2011,
  \href{http://dx.doi.org/10.1088/1742-6596/271/1/012049}{Journal of Physics:
  Conference Series, 271, 2049}

\bibitem[{Garcia {et~al.}(2001)Garcia, Regulo, Turck-Chieze, Bertello,
  Kosovichev, Brun, Couvidat, Henney, Lazrek, Ulrich, \&
  Varadi}]{2001SoPh..200..361G}
Garcia, R.~A., Regulo, C., Turck-Chieze, S., {et~al.} 2001,
  \href{http://adsabs.harvard.edu/cgi-bin/nph-data_query?bibcode=2001SoPh..200..361G&link_type=ABSTRACT}{Solar
  Physics, 200, 361}

\bibitem[{Goldreich \& Keeley(1977)}]{1977ApJ...211..934G}
Goldreich, P., \& Keeley, D.~A. 1977,
  \href{http://dx.doi.org/10.1086/155005}{Astrophysical Journal, 211, 934}

\bibitem[{Goldreich {et~al.}(1994)Goldreich, Murray, \&
  Kumar}]{1994ApJ...424..466G}
Goldreich, P., Murray, N., \& Kumar, P. 1994,
  \href{http://dx.doi.org/10.1086/173904}{Astrophysical Journal, 424, 466}

\bibitem[{Gottardi(2007)}]{2007PhRvD..75b2002G}
Gottardi, L. 2007, \href{http://dx.doi.org/10.1103/PhysRevD.75.022002}{Physical
  Review D, 75, 22002}

\bibitem[{Grigahc{\`e}ne {et~al.}(2005)Grigahc{\`e}ne, Dupret, Gabriel,
  Garrido, \& Scuflaire}]{2005A&A...434.1055G}
Grigahc{\`e}ne, A., Dupret, M.~A., Gabriel, M., Garrido, R., \& Scuflaire, R.
  2005, \href{http://dx.doi.org/10.1051/0004-6361:20041816}{Astronomy and
  Astrophysics, 434, 1055}

\bibitem[{Haxton {et~al.}(2013)Haxton, Hamish~Robertson, \&
  Serenelli}]{2013ARA&A..51...21H}
Haxton, W.~C., Hamish~Robertson, R.~G., \& Serenelli, A.~M. 2013,
  \href{http://dx.doi.org/10.1146/annurev-astro-081811-125539}{Annual Review of
  Astronomy and Astrophysics, 51, 21}

\bibitem[{Houdek {et~al.}(1999)Houdek, Balmforth, Christensen-Dalsgaard, \&
  Gough}]{1999A&A...351..582H}
Houdek, G., Balmforth, N.~J., Christensen-Dalsgaard, J., \& Gough, D.~O. 1999,
\href{http://adsabs.harvard.edu/abs/1999A&A...351..582H}{Astronomy and Astrophysics, 351, 582}

\bibitem[{Houdek \& Gough(2002)}]{2002MNRAS.336L..65H}
Houdek, G., \& Gough, D.~O. 2002,
  \href{http://dx.doi.org/10.1046/j.1365-8711.2002.06024.x}{Monthly Notice of
  the Royal Astronomical Society, 336, L65}

\bibitem[{Howe(2009)}]{2009LRSP....6....1H}
Howe, R. 2009, \href{http://dx.doi.org/10.12942/lrsp-2009-1}{Living Reviews in
  Solar Physics, 6, 1}

\bibitem[{Jimenez \& Garcia(2009)}]{2009ApJS..184..288J}
Jimenez, A., \& Garcia, R.~A. 2009,
  \href{http://dx.doi.org/10.1088/0067-0049/184/2/288}{The Astrophysical
  Journal Supplement, 184, 288}

\bibitem[{Khosroshahi \& Sobouti(1997)}]{1997A&A...321.1024K}
Khosroshahi, H.~G., \& Sobouti, Y. 1997,
  \href{http://adsabs.harvard.edu/abs/1997A&A...321.1024K}{Astronomy and
  Astrophysics, 321, 1024}

\bibitem[{Kosovichev(1995)}]{1995ESASP.376a.165K}
Kosovichev, A.~G. 1995, Helioseismology. ESA SP, Proceedings of the 4th Soho Workshop, held Pacific Grove, California, USA, 2-6 April 1995, Paris: European Space Agency (ESA), 
edited by J.T. Hoeksema, V. Domingo, B. Fleck, and Bruce Battrick, Invited Reviews and Working Group Reports,
  \href{http://adsabs.harvard.edu/abs/1995ESASP.376a.165K}{Helioseismology. ESA
  SP, 376, 165}

\bibitem[{Libbrecht(1988)}]{1988ApJ...334..510L}
Libbrecht, K.~G. 1988, \href{http://dx.doi.org/10.1086/166855}{Astrophysical
  Journal, 334, 510}

\bibitem[{Lopes \& Silk(2013)}]{2013MNRAS.435.2109L}
Lopes, I., \& Silk, J. 2013,
  \href{http://dx.doi.org/10.1093/mnras/stt1427}{Monthly Notices of the Royal
  Astronomical Society, 435, 2109}

\bibitem[{Lopes(2001)}]{2001A&A...373..916L}
Lopes, I.~P. 2001,
  \href{http://dx.doi.org/10.1051/0004-6361:20010130}{Astronomy and
  Astrophysics, 373, 916}

\bibitem[{Lopes \& Gough(2001)}]{2001MNRAS.322..473L}
Lopes, I.~P., \& Gough, D. 2001,
  \href{http://dx.doi.org/10.1046/j.1365-8711.2001.03940.x}{Monthly Notices of
  the Royal Astronomical Society, 322, 473}

\bibitem[{Lynden-Bell \& Rees(1971)}]{1971MNRAS.152..461L}
Lynden-Bell, D., \& Rees, M.~J. 1971,
  \href{http://adsabs.harvard.edu/abs/1971MNRAS.152..461L}{Monthly Notices of
  the Royal Astronomical Society, 152, 461}

\bibitem[{Maggiore(2008)}]{Maggiore:2008tka}
Maggiore, M. 2008, {Gravitational Waves}, Volume 1: Theory and Experiments
  (Oxford University Press)

\bibitem[{McKernan {et~al.}(2014)McKernan, Ford, Kocsis, \&
  Haiman}]{2014arXiv1405.1414M}
McKernan, B., Ford, K. E.~S., Kocsis, B., \& Haiman, Z. 2014,
  \href{http://adsabs.harvard.edu/abs/2014arXiv1405.1414M}{MNRAS Letters accepted, eprint arXiv:1405.1414}

\bibitem[{Michel {et~al.}(2008)Michel, Baglin, Auvergne, Catala, \&
  Samadi}]{2008Sci...322..558M}
Michel, E., Baglin, A., Auvergne, M., Catala, C., \& Samadi, R. 2008,
\href{http://dx.doi.org/10.1126/science.1163004}{Science, Volume 322, Issue 5901,pp. 558.}

\bibitem[{Misner {et~al.}(1973)Misner, Thorne, \&
  Wheeler}]{1973grav.book.....M}
Misner, C.~W., Thorne, K.~S., \& Wheeler, J.~A. 1973,
  \href{http://adsabs.harvard.edu/abs/1973grav.book.....M}{San Francisco: W.H.
  Freeman and Co.}

\bibitem[{Moore et al. (2014)}]{Moore:wk}
Moore, C. J.; Cole, R. H.; Berry, C. P. L.
\href{http://adsabs.harvard.edu/abs/2014arXiv1408.0740M}{eprint arXiv:1408.0740}

\bibitem[{Morel(1997)}]{1997A&AS..124..597M}
Morel, P. 1997, \href{http://dx.doi.org/10.1051/aas:1997209}{A {\&} A
  Supplement series, 124, 597}

\bibitem[{Nelemans {et~al.}(2004)Nelemans, Yungelson, \&
  Portegies~Zwart}]{2004MNRAS.349..181N}
Nelemans, G., Yungelson, L.~R., \& Portegies~Zwart, S.~F. 2004,
  \href{http://dx.doi.org/10.1111/j.1365-2966.2004.07479.x}{Monthly Notices of
  the Royal Astronomical Society, 349, 181}

\bibitem[{Provost {et~al.}(2000)Provost, Berthomieu, \&
  Morel}]{2000A&A...353..775P}
Provost, J., Berthomieu, G., \& Morel, P. 2000,
  \href{http://adsabs.harvard.edu/cgi-bin/nph-data_query?bibcode=2000A%26A...353..775P&link_type=ABSTRACT}{Astronomy
  and Astrophysics, 353, 775}

\bibitem[{Rathore {et~al.}(2004)Rathore, Blandford, \&
  Broderick}]{2004AAS...205.7605R} Rathore, Y., Blandford, R.~D., \& Broderick, A.~E. 2004,
American Astronomical Society Meeting 205, \#76.05; \href{http://adsabs.harvard.edu/abs/2004AAS...205.7605R}{Bulletin of American Astronomical Society, Vol. 36, p.1473}

\bibitem[{Roelofs {et~al.}(2007)Roelofs, Groot, Benedict, McArthur, Steeghs,
  Morales-Rueda, Marsh, \& Nelemans}]{2007ApJ...666.1174R}
Roelofs, G. H.~A., Groot, P.~J., Benedict, G.~F., {et~al.} 2007,
  \href{http://dx.doi.org/10.1086/520491}{The Astrophysical Journal, 666, 1174}

\bibitem[{Samadi \& Goupil(2001)}]{2001A&A...370..136S}
Samadi, R., \& Goupil, M.~J. 2001,
  \href{http://dx.doi.org/10.1051/0004-6361:20010212}{Astronomy and
  Astrophysics, 370, 136}

\bibitem[{Samadi {et~al.}(2001)Samadi, Goupil, \&
  Lebreton}]{2001A&A...370..147S}
Samadi, R., Goupil, M.~J., \& Lebreton, Y. 2001,
  \href{http://dx.doi.org/10.1051/0004-6361:20010213}{Astronomy and
  Astrophysics, 370, 147}

\bibitem[{Sathyaprakash \& Schutz(2009)}]{2009LRR....12....2S}
Sathyaprakash, B.~S., \& Schutz, B.~F. 2009,
  \href{http://dx.doi.org/10.12942/lrr-2009-2}{Living Reviews in Relativity,
  12, 2}

\bibitem[{Schutz(2009)}]{2009fcgr.book.....S}
Schutz, B. 2009, {A First Course in General Relativity} (A First Course in
  General Relativity by Bernard Schutz. Cambridge University Press)

\bibitem[{Siegel \& Roth(2010)}]{2010MNRAS.408.1742S}
Siegel, D.~M., \& Roth, M. 2010,
  \href{http://dx.doi.org/10.1111/j.1365-2966.2010.17240.x}{Monthly Notices of
  the Royal Astronomical Society, 408, 1742}

\bibitem[{Siegel \& Roth(2011)}]{2011ApJ...729..137S}
---. 2011, \href{http://dx.doi.org/10.1088/0004-637X/729/2/137}{The
  Astrophysical Journal, 729, 137}

\bibitem[{Siegel \& Roth(2014)}]{2014ApJ...784...88S}
---. 2014, \href{http://dx.doi.org/10.1088/0004-637X/784/2/88}{The
  Astrophysical Journal, 784, 88}

\bibitem[{Turck-Chieze \& Lopes(1993)}]{1993ApJ...408..347T}
Turck-Chieze, S., \& Lopes, I. 1993,
  \href{http://dx.doi.org/10.1086/172592}{Astrophysical Journal, 408, 347}

\bibitem[{Turck-Chieze \& Lopes(2012)}]{2012RAA....12.1107T}
---. 2012, \href{http://dx.doi.org/10.1088/1674-4527/12/8/011}{Research in
  Astronomy and Astrophysics, 12, 1107}

\bibitem[{Turck-Chieze {et~al.}(2004)Turck-Chieze, Garcia, Couvidat, Ulrich,
  Bertello, Varadi, Kosovichev, Gabriel, Berthomieu, Brun, Lopes, Palle,
  Provost, Robillot, \& Roca-Cortes}]{2004ApJ...604..455T}
Turck-Chieze, S., Garcia, R.~A., Couvidat, S., {et~al.} 2004,
  \href{http://dx.doi.org/10.1086/381743}{The Astrophysical Journal, 604, 455}

\bibitem[{Unno {et~al.}(1989)Unno, Osaki, Ando, Saio, \&
  Shibahashi}]{1989nos..book.....U}
Unno, W., Osaki, Y., Ando, H., Saio, H., \& Shibahashi, H. 1989,
  \href{http://adsabs.harvard.edu/abs/1989nos..book.....U}{Nonradial
  oscillations of stars, Tokyo: University of Tokyo Press, 1989, 2nd ed., -1}

\bibitem[{Yu \& Jeffery(2010)}]{2010A&A...521A..85Y}
Yu, S., \& Jeffery, C.~S. 2010,
  \href{http://dx.doi.org/10.1051/0004-6361/201014827}{Astronomy and
  Astrophysics, 521, 85}

\bibitem[{Zimmerman \& Hellings(1980)}]{1980ApJ...241..475Z}
Zimmerman, R.~L., \& Hellings, R.~W. 1980,
  \href{http://dx.doi.org/10.1086/158362}{Astrophysical Journal, 241, 475}

  
\end{thebibliography}

\end{document}